\documentclass[11pt]{article}
\usepackage{moriond,epsfig}

\def\be{\begin{equation}}
\def\ee{\end{equation}}
\def\bea{\begin{eqnarray}}
\def\eea{\end{eqnarray}}

\begin{document}
\vspace*{4cm}
\title{A Lepton Universality Test at CERN NA62 Experiment}

\author{Evgueni Goudzovski}

\address{School of Physics and Astronomy, University of
Birmingham,\\
Edgbaston, Birmingham, B15 2TT, United Kingdom}

\maketitle\abstracts{The NA62 experiment at CERN collected a large
sample of $K^+\to e^+\nu$ decays during a dedicated run in 2007,
aiming at a precise test of lepton universality by measurement of
the helicity suppressed ratio $R_K = \Gamma(K^+\to
e^+\nu)/\Gamma(K^+\to\mu^+\nu)$. The preliminary result of the
analysis of a partial data sample of 51089 $K^+\to e^+\nu$
candidates is $R_K=(2.500\pm0.016)\times10^{-5}$, which is
consistent with the Standard Model expectation.}

\section*{Introduction}

Due to the $V-A$ structure of the weak interactions, the Standard
Model (SM) rates of the leptonic meson decays $P^+\to\ell^+\nu$ are
helicity suppressed. Within the two-Higgs doublet models (2HDM),
which is a wide class of models including the minimum supersymmetric
(SUSY) one, the charged Higgs boson ($H^\pm$) exchange induces a
tree-level contribution to (semi)leptonic decays proportional to the
Yukawa couplings of quarks and leptons~\cite{ho93}. In
$P^+\to\ell^+\nu$ decays, the $H^\pm$ exchange can compete with the
$W^\pm$ exchange, thanks to the above suppression.

At tree level, the $H^\pm$ exchange contribution to
$P^+\to\ell^+\nu$ decay widths (with $P=\pi, K, B$) is lepton
flavour independent, and is approximately described by~\cite{is06}
\begin{equation}
\frac{\Delta\Gamma(P^+\to\ell^+\nu)}{\Gamma^{\rm
SM}(P^+\to\ell^+\nu)} \approx -2\left(\frac{M_P}{M_H}\right)^2
\frac{\tan^2\beta}{1+\varepsilon_0\tan\beta}. \label{eq:tree}
\end{equation}
Here $M_H$ is the charged Higgs boson mass, $\tan\beta$ is the ratio
of vacuum expectation values of the two Higgs doublets, a
fundamental parameter controlling the charged Higgs couplings, and
$\varepsilon_0\sim 10^{-2}$ is an effective coupling. For a
reasonable choice of the parameters ($\tan\beta=40$, $M_H=500~{\rm
GeV}/c^2$), one expects $\sim 30\%$ relative suppression of
$B^+\to\ell^+\nu$ decays, and $\sim 0.3\%$ suppression of
$K^+\to\ell^+\nu$ decays. However the searches for new physics in
these decay rates are hindered by the uncertainties of their SM
predictions.

On the other hand, the ratio of kaon leptonic decay rates
$R_K=\Gamma(K_{e2})/\Gamma(K_{\mu2})$, where the notation $K_{\ell
2}$ is adopted for the $K^+\to\ell^+\nu$ decays, has been calculated
with an excellent accuracy within the SM~\cite{ci07}:
\begin{equation}
R_K^\mathrm{SM} = \left(m_e/m_\mu\right)^2
\left(\frac{m_K^2-m_e^2}{m_K^2-m_\mu^2}\right)^2 (1 + \delta
R_{\mathrm{QED}})= (2.477 \pm 0.001)\times 10^{-5}.
\end{equation}
Here $\delta R_{\mathrm{QED}}=(-3.78\pm0.04)\%$ is a correction due
to the inner bremsstrahlung (IB) $K_{\ell 2\gamma}$ process. The
ratio $R_K$ is sensitive to lepton flavour universality violation
(LFV) effe
cts originating at one-loop level from $H^\pm$ exchange in
2HDM~\cite{ma06,ma08}, and the mixing effects in the right-handed
slepton sector, providing a unique probe into this aspect of
supersymmetric flavour physics~\cite{el09}. $R_K$ receives the
following leading-order contribution due to LFV coupling of the
Higgs boson:
\begin{equation}
\frac{\Delta R_K}{R_K^\mathrm{SM}}=\left(\frac{M_K}{M_H}\right)^4
\left(\frac{M_\tau}{M_e}\right)^2 |\Delta _R^{31}|^2\tan^6\beta,
\end{equation}
where $|\Delta_{R}^{31}|$ is the mixing parameter between the
superpartners of the right-handed leptons, which can reach $\sim
10^{-3}$. This can enhance $R_K$ by ${\cal O}(1\%)$ relative, with
no contradiction to presently known experimental constraints
(including upper bounds on the LFV $\tau\to eX$ decays with
$X=\eta,\gamma,\mu\mu$).

The current world average (including only final results, and thus
ignoring the preliminary NA48/2 ones) is $R_K^{\rm
WA}=(2.490\pm0.030)\times 10^{-5}$, dominated by a recent
measurement by the KLOE collaboration~\cite{am09}. The NA62
experiment at CERN collected a dedicated data sample in 2007--08,
aiming at a measurement of $R_K$ with a $0.4\%$ precision. The
preliminary result obtained with a partial data sample is presented
here.

\section{Beam, detector and data taking}

The beam line and setup of the NA48/2 experiment~\cite{fa07} were
used for the NA62 2007--08 data taking. Experimental conditions and
trigger logic were optimized for the $K_{e2}/K_{\mu2}$ measurement.

The beam line is capable of delivering simultaneous unseparated
$K^+$ and $K^-$ beams derived from 400 GeV/$c$ primary protons
extracted from the CERN SPS. Most of the data, including the sample
used for the present analysis, were collected with the $K^+$ beam
only, as the muon sweeping system provides better suppression of the
positive beam halo component. A narrow momentum band of
$(74.0\pm1.6)$ GeV/$c$ was used to minimize the corresponding
contribution to resolution in kinematical variables.

The fiducial decay region is contained in a 114 m long cylindrical
vacuum tank. With $1.8\times 10^{12}$ primary protons incident on
the target per SPS pulse of $4.8$~s duration, the beam flux at the
entrance to the decay volume is $2.5\times 10^7$ particles per
pulse. The fractions of $K^+$, $\pi^+$, $p$, $e^+$ and $\mu^+$ in
the beam are 0.05, 0.63, 0.21, 0.10 and 0.01, respectively. The
fraction of beam kaons decaying in the vacuum tank at nominal
momentum is $18\%$. The transverse size of the beam within the decay
volume is $\delta x = \delta y = 7$~mm (rms), and its angular
divergence is negligible.

Among the subdetectors located downstream the decay volume, a
magnetic spectrometer, a plastic scintillator hodoscope (HOD) and a
liquid krypton electromagnetic calorimeter (LKr) are principal for
the measurement. The spectrometer, used to detect charged products
of kaon decays, is composed of four drift chambers (DCHs) and a
dipole magnet. The HOD producing fast trigger signals consists of
two planes of strip-shaped counters. The LKr, used for particle
identification and as a veto, is an almost homogeneous ionization
chamber, $27X_0$ deep, segmented transversally into 13,248 cells
(2$\times$2 cm$^2$ each), and with no longitudinal segmentation. A
beam pipe traversing the centres of the detectors allows undecayed
beam particles and muons from decays of beam pions to continue their
path in vacuum.

A minimum bias trigger configuration is employed, resulting in high
efficiency with relatively low purity. The $K_{e2}$ trigger
condition consists of coincidence of hits in the HOD planes (the so
called $Q_1$ signal) with 10 GeV LKr energy deposition. The
$K_{\mu2}$ trigger condition consists of the $Q_1$ signal alone
downscaled by a factor of 150. Loose lower and upper limits on DCH
activity are also applied.

The main data taking took place during four months starting in June
2007. Two additional weeks of data taking allocated in September
2008 were used to collect special data samples for studies of
systematic effects. The present analysis is based on $\sim 40\%$ of
the data sample.

\section{Analysis strategy and event selection}

The analysis strategy is based on counting the numbers of
reconstructed $K_{e2}$ and $K_{\mu2}$ candidates collected
concurrently. Consequently the result does not rely on kaon flux
measurement, and several systematic effects (e.g. due to
reconstruction and trigger efficiencies, time-dependent effects)
cancel to first order.

To take into account the significant dependence of signal acceptance
and background level on lepton momentum, the measurement is
performed independently in bins of this observable: 10 bins covering
a lepton momentum range of $[15; 65]$~GeV/$c$ are used. The ratio
$R_K$ in each bin is computed as
\begin{equation}
R_K = \frac{1}{D}\cdot \frac{N(K_{e2})-N_{\rm
B}(K_{e2})}{N(K_{\mu2}) - N_{\rm B}(K_{\mu2})}\cdot
\frac{A(K_{\mu2})}{A(K_{e2})} \cdot
\frac{f_\mu\times\epsilon(K_{\mu2})}
{f_e\times\epsilon(K_{e2})}\cdot\frac{1}{f_\mathrm{LKr}},
\label{eq:rkcomp}
\end{equation}
where $N(K_{\ell 2})$ are the numbers of selected $K_{\ell 2}$
candidates $(\ell=e,\mu)$, $N_{\rm B}(K_{\ell 2})$ are numbers of
background events, $A(K_{\mu 2})/A(K_{e2})$ is the geometric
acceptance correction, $f_\ell$ are efficiencies of $e$/$\mu$
identification, $\epsilon(K_{\ell 2})$ are trigger efficiencies,
$f_\mathrm{LKr}$ is the global efficiency of the LKr readout, and
$D=150$ is the downscaling factor of the $K_{\mu2}$ trigger.

A detailed Monte Carlo (MC) simulation including beam line optics,
full detector geometry and material description, stray magnetic
fields, local inefficiencies of DCH wires, and time variations of
the above throughout the running period, is used to evaluate the
acceptance correction $A(K_{\mu2})/A(K_{e2})$ and the geometric
parts of the acceptances for background processes entering the
computation of $N_B(K_{\ell 2})$. The $K_{\ell 2 (\gamma)}$
processes are simulated in one-photon approximation~\cite{ci07}; the
resummation of leading logarithms~\cite{ga06} is neglected at this
stage. Simulations are used to a limited extent only: particle
identification, trigger and readout efficiencies are measured
directly.

Due to topological similarity of $K_{e2}$ and $K_{\mu2}$ decays, a
large part of the selection conditions is common for both decays:
(1) exactly one reconstructed particle of positive electric charge;
(2) its momentum $15~{\textrm{GeV}}/c<p<65~{\textrm{GeV}}/c$ (the
lower limit is due to the 10 GeV LKr energy deposit trigger
requirement in $K_{e2}$ trigger); (3) extrapolated track impact
points in DCH, LKr and HOD are within their geometrical acceptances;
(4) no LKr energy deposition clusters with energy $E>2$~GeV and not
associated to the track to suppress background from other kaon
decays; (5) distance between the charged track and the nominal kaon
beam axis ${\rm CDA}<3$~cm, decay vertex longitudinal position
within the nominal decay volume (the latter condition is optimized
in each lepton momentum bin).

The following two principal selection criteria are different for the
$K_{e2}$ and $K_{\mu2}$ decays. $K_{\ell 2}$ kinematic
identification is based on the reconstructed squared missing mass
assuming the track to be a positron or a muon:
$M_{\mathrm{miss}}^2(\ell) = (P_K - P_\ell)^2$, where $P_K$ and
$P_\ell$ ($\ell = e,\mu$) are the four-momenta of the kaon (average
beam momentum assumed) and the lepton (positron or muon mass
assumed). A selection condition $|M_{\mathrm{miss}}^2(e)|<M_0^2$ is
applied to select $K_{e2}$ candidates, and
$|M_{\mathrm{miss}}^2(\mu)|<M_0^2$ for $K_{\mu2}$ ones, where
$M_0^2$ varies from 0.009 to 0.013~$(\mathrm{GeV}/c^2)^2$ among
lepton momentum bins depending on $M_{\mathrm{miss}}^2$ resolution.
Particle identification is based on the ratio $E/p$ of track energy
deposit in the LKr calorimeter to its momentum measured by the
spectrometer. Particles with $0.95<E/p<1.1$ ($E/p<0.85$) are
identified as positrons (muons).

\section{Backgrounds}

\begin{figure}[tb]
\begin{center}
\resizebox{0.51\textwidth}{!}{\includegraphics{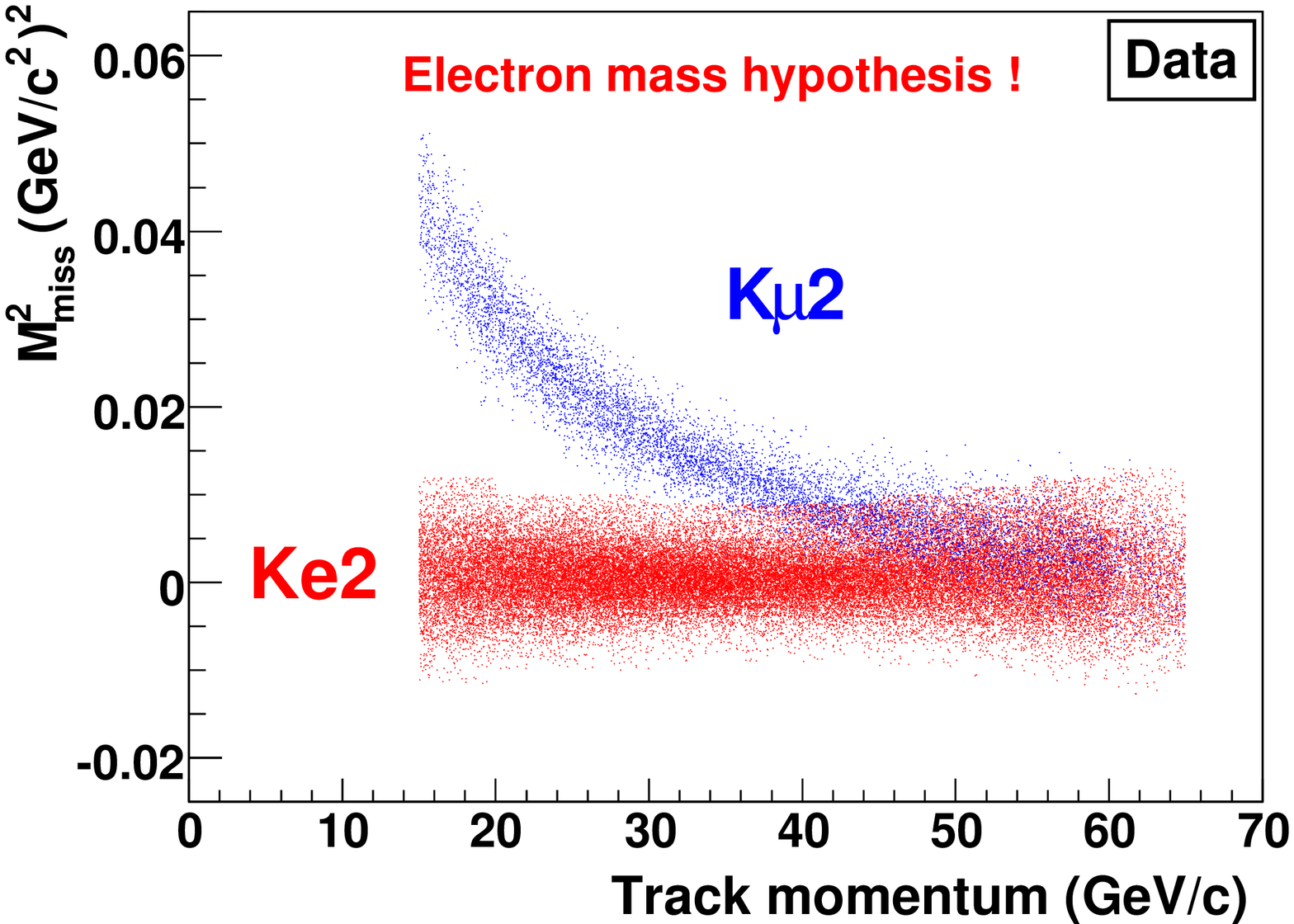}}%
\resizebox{0.49\textwidth}{!}{\includegraphics{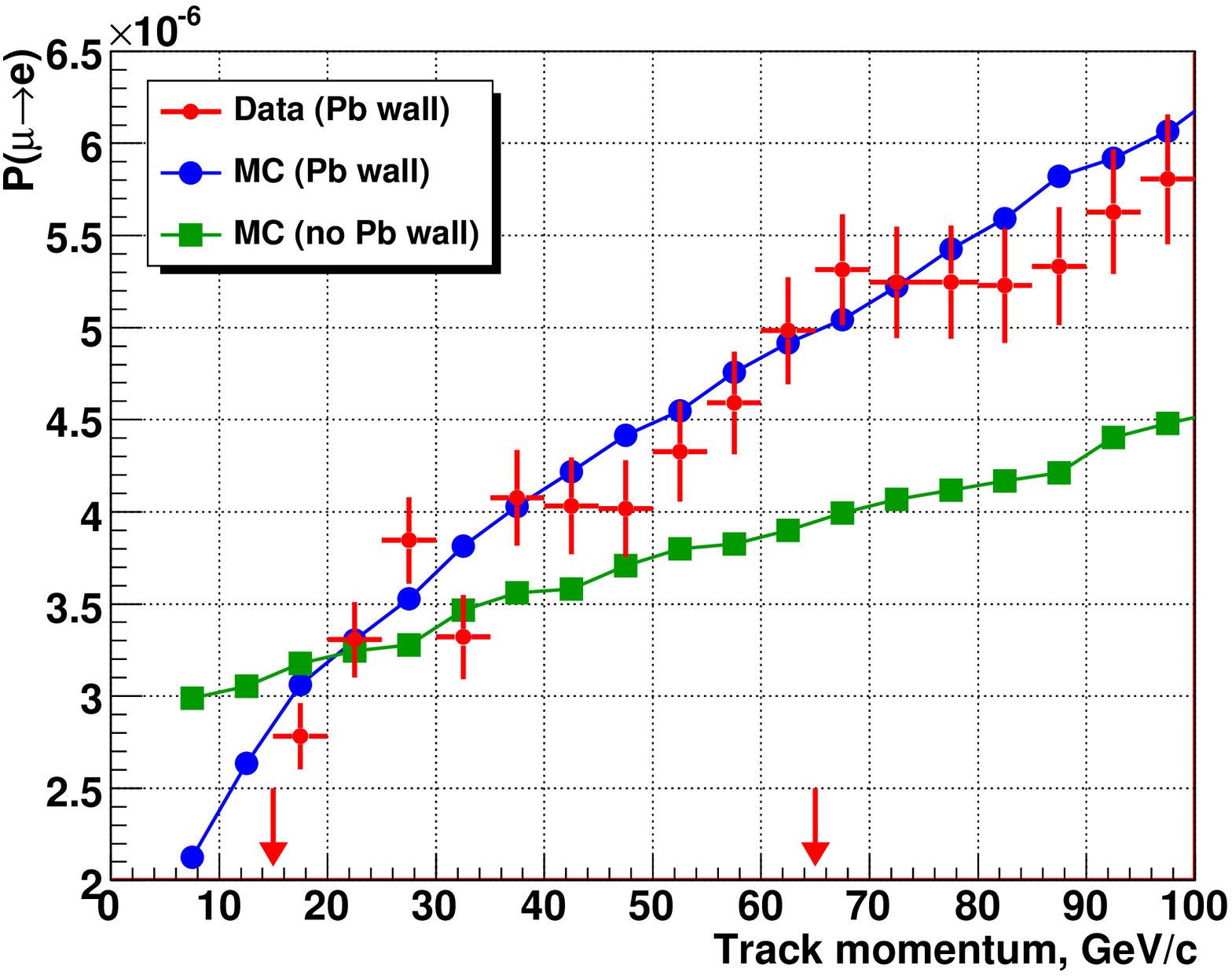}}
\end{center}
\vspace{-3mm} \caption{(a) Missing mass squared in positron
hypothesis $M_{\rm miss}^2(e)$ vs lepton momentum for reconstructed
$K_{e2}$ and $K_{\mu2}$ decays: kinematic separation of $K_{e2}$ and
$K_{\mu2}$ decays is possible at low lepton momentum only. (b)
Measured and simulated probability of muon identification as
electron/positron $P(\mu\to e)$ vs its momentum: data with the Pb
wall, MC simulations with and without the Pb wall (the signal region
is marked with arrows).} \label{fig:pbwall}
\end{figure}

\boldmath {\bf $K_{\mu2}$ decay} \unboldmath with a mis-identified
muon is the main background source in the $K_{e2}$ sample.
Sufficient kinematic separation of $K_{e2}$ and $K_{\mu2}$ decays is
not achievable at high lepton momentum ($p>30$~GeV/$c$), as shown in
Fig.~\ref{fig:pbwall}a. The probability of muon identification as
positron in that momentum range ($E/p>0.95$ due to `catastrophic'
bremsstrahlung in or in front of the LKr) is $P(\mu\to
e)\sim4\times10^{-6}$, which is non-negligible compared to $R_K^{\rm
SM}=2.477\times 10^{-5}$. A direct measurement of $P(\mu\to e)$ to
$\sim 10^{-2}$ relative precision is necessary to validate the
theoretical calculation of the bremsstrahlung
cross-section~\cite{ke97} in the high $\gamma$ energy range used to
evaluate the $K_{\mu2}$ background.

The available muon samples are typically affected by $\sim10^{-4}$
electron/positron contamination due to $\mu\to e$ decays in flight,
which obstructs the $P(\mu\to e)$ measurements. In order to obtain
sufficiently pure muon samples, a $9.2X_0$ thick lead (Pb) wall
covering $\sim 20\%$ of the geometric acceptance was installed in
front of the LKr calorimeter (between the two HOD planes) during a
period of the data taking. In the samples of tracks traversing the
Pb and having $E/p>0.95$, the electron component is suppressed to a
level of $\sim 10^{-7}$ by energy losses in Pb.

The momentum dependence of $P(\mu\to e)$ for muons traversing the Pb
has been measured with a data sample collected during a special muon
($\mu^\pm$) run of 20h duration, and compared to the results of a
dedicated Geant4-based MC simulation of the region downstream the
spectrometer involving standard energy loss processes and
bremsstrahlung~\cite{ke97}. The data/MC comparison
(Fig.~\ref{fig:pbwall}b) shows good agreement in a wide momentum
range within statistical errors, which validates the cross-section
calculation at the corresponding precision level. The simulation
shows that the Pb wall modifies $P(\mu\to e)$ via two principal
mechanisms: 1) muon energy loss in the Pb by ionization decreasing
$P(\mu\to e)$ and dominating at low momentum; 2) bremsstrahlung in
Pb increasing $P(\mu\to e)$ and dominating at high momentum.

To estimate the $K_{\mu2}$ background contamination, the kinematic
suppression factor is computed with the standard setup simulation,
while the validated simulation of muon interaction in the LKr is
employed to account for $P(\mu\to e)$ suppression. Uncertainty of
the background estimate is due to the limited size of the data
sample used to validate the simulation.

\boldmath {\bf $K_{\mu2}$ decay followed by $\mu\to e$ decay}
\unboldmath contributes significantly to the background. However
energetic forward daughter positrons compatible to $K_{e2}$ topology
are suppressed due to muon polarization~\cite{mi50}.

\boldmath {\bf $K_{e2\gamma}$ (SD) decay}, \unboldmath a background
by $R_K$ definition, has a rate similar to that of $K_{e2}$: the
world average~\cite{pdg} is
$\mathrm{BR}=(1.52\pm0.23)\times10^{-5}$. Theoretical rate
calculations depend on the form factor model, and have a similar
precision. Energetic positrons ($E_e^*>230$~MeV in $K^+$ frame) with
$\gamma$ escaping detector acceptance contribute to the background.
MC background estimation has a 15\% uncertainty due to limited
knowledge of the process. A recent measurement by KLOE~\cite{am09},
published after announcement of the NA62 preliminary result, is not
used.

{\bf Beam halo} background in the $K_{e2}$ sample induced by halo
muons (undergoing $\mu\to e$ decay in flight or mis-identified) is
measured directly by reconstructing $K^+_{e2}$ candidates from a
control $K^-$ data sample collected with the $K^+$ beam dumped.
Background rate and kinematical distribution are qualitatively
reproduced by a halo simulation. The uncertainty is due to the
limited size of the $K^-$ sample. Beam halo is the only significant
background source in the $K_{\mu2}$ sample, measured to be $0.25\%$
(with a negligible uncertainty) with the same technique as for
$K_{e2}$ decays.

\begin{figure}[tb]
\begin{center}
\resizebox*{0.47\textwidth}{!}{\includegraphics{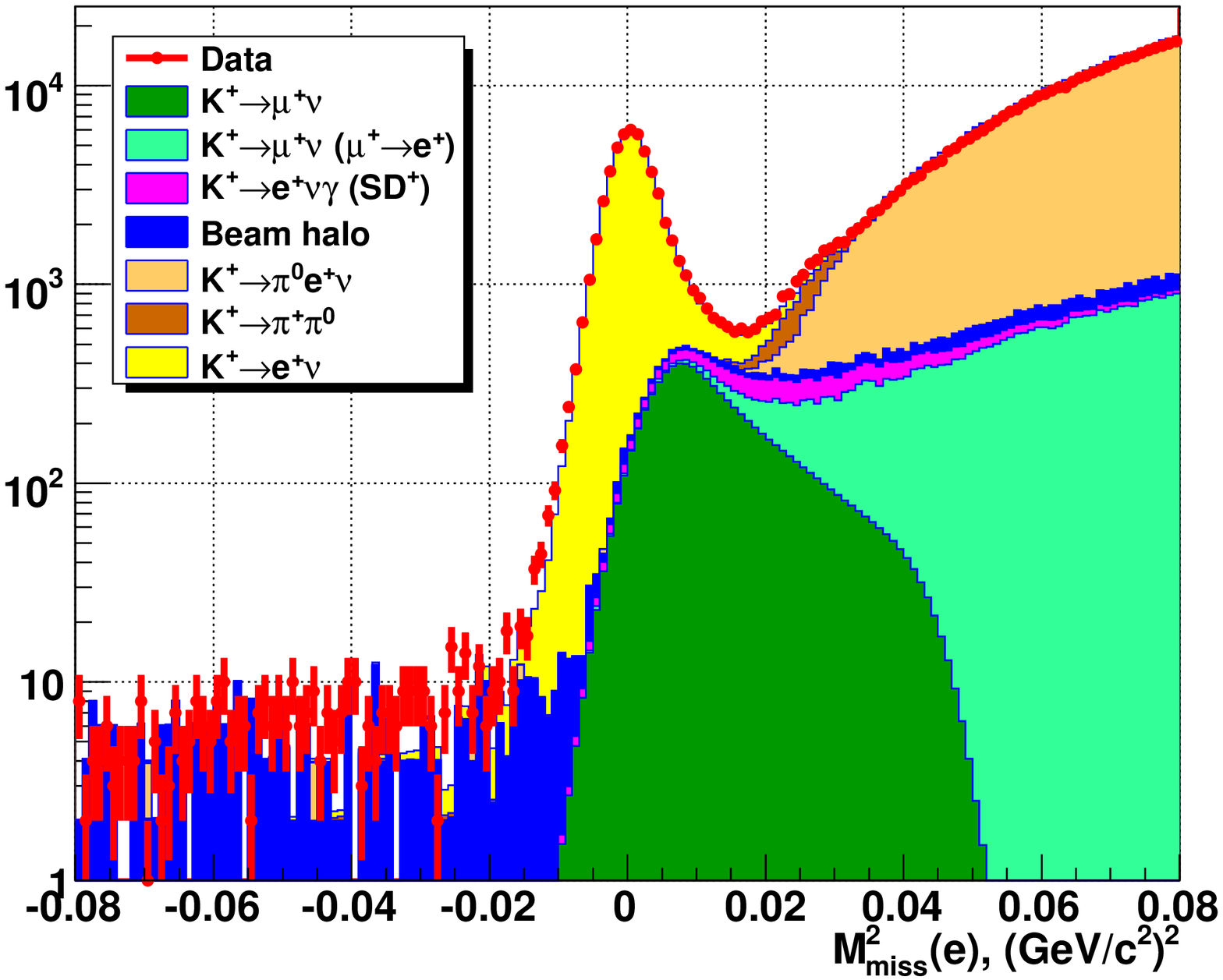}}~~
\resizebox*{0.45\textwidth}{!}{\includegraphics{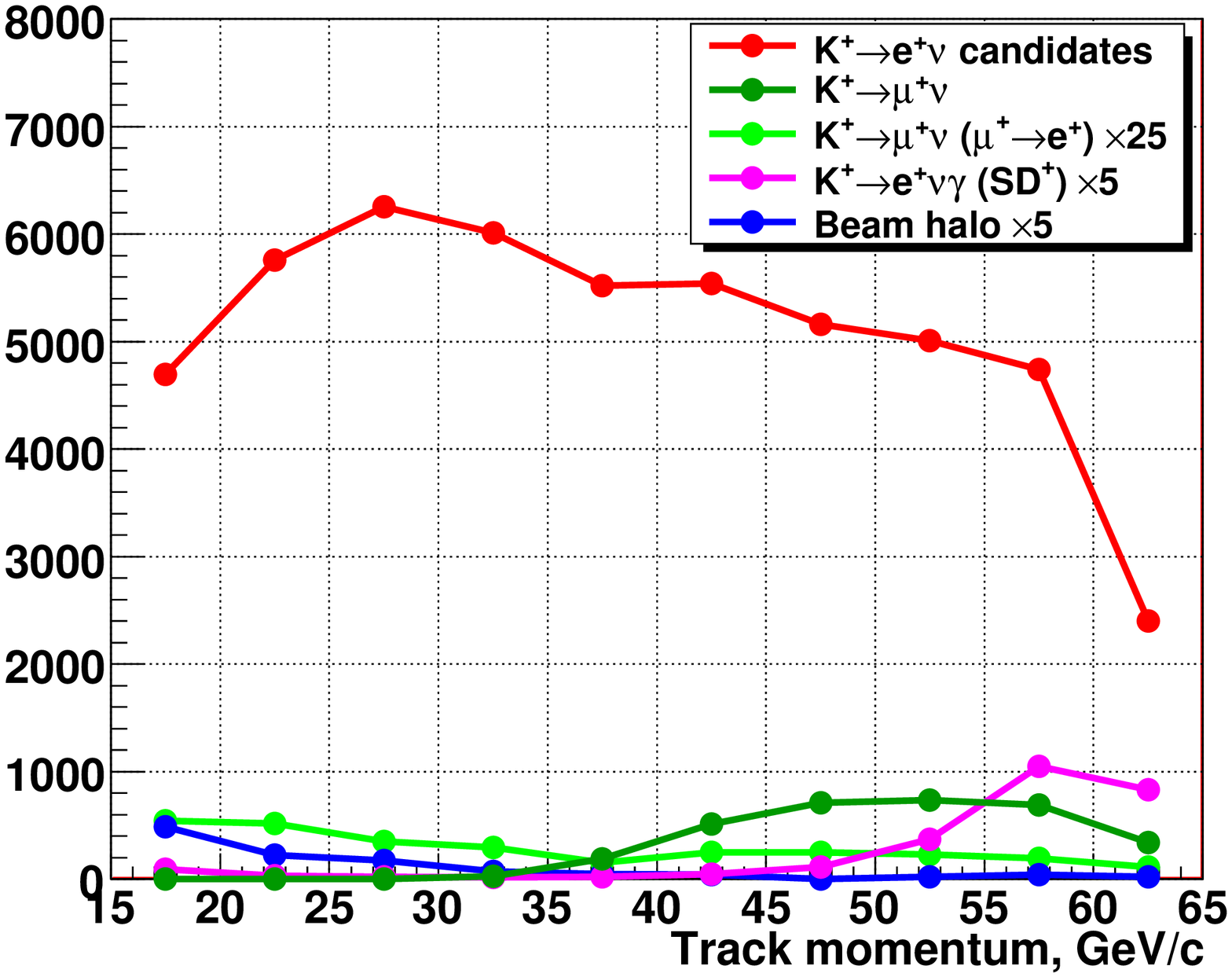}}
\end{center}
\vspace{-3mm} \caption{(a) Reconstructed squared missing mass
distribution $M_{\mathrm{miss}}^2(e)$ for the $K_{e2}$ candidates:
data (dots) presented as sum of MC signal and background
contributions (filled areas). (b) Numbers of $K_{e2}$ candidates and
background events in lepton momentum bins.} \label{fig:mmiss}
\end{figure}

\begin{table}
\begin{center}
\begin{tabular}{lc|lc|lc}
\hline Source & $N_B/N_{tot}$ & Source & $N_B/N_{tot}$ & Source & $N_B/N_{tot}$\\
\hline
$K_{\mu2}$               & $(6.28\pm0.17)\%$ & $K_{e2\gamma}$~(SD) & $(1.02\pm0.15)\%$ & $K_{e3}$   & $0.03\%$\\
$K_{\mu2}~(\mu\to e)$    & $(0.23\pm0.01)\%$ & Beam halo  & $(1.45\pm0.04)\%$          & $K_{2\pi}$ & $0.03\%$\\
\hline
\multicolumn{6}{c}{Total background: $(8.03\pm0.23)\%$}\\
\hline
\end{tabular}
\end{center}
\caption{Summary of the background sources in the $K_{e2}$ sample.}
\label{tab:bkg}
\end{table}

The number of $K_{\ell 2}$ candidates is $N(K_{e2})=51,089$ (about
four times the statistics collected by KLOE~\cite{am09}) and
$N(K_{\mu2})=15.56\times 10^6$. The $M_{\mathrm{miss}}^2(e)$
distributions of data events and backgrounds are presented in
Fig.~\ref{fig:mmiss}a. Backgrounds integrated over lepton momentum
are summarized in Table~\ref{tab:bkg}; their distributions over
lepton momentum are presented in Fig.~\ref{fig:mmiss}b.

\section{Systematic uncertainties}

{\bf Positron identification efficiency} is measured directly as a
function of momentum and LKr impact point using pure samples of
electrons and positrons obtained by kinematic selection of
$K^+\to\pi^0 e^+\nu$ decays collected concurrently with the $K_{e2}$
sample, and $K^0_L\to\pi^\pm e^\mp\nu$ decays from a special $K^0_L$
run of 15 hours duration. The $K^+$ and $K^0_L$ measurements are in
good agreement. The measured $f_e$ averaged over the $K_{e2}$ sample
is $(99.20\pm0.05)\%$. Muon identification inefficiency is
negligible.

{\bf The geometric acceptance correction} $A(K_{\mu2})/A(K_{e2})$ is
strongly affected by the radiative $K_{e2\gamma}$ (IB) decays. A
conservative systematic uncertainty is attributed to approximations
used in the $K_{e2\gamma}$ IB simulation. The resummation of leading
logarithms~\cite{ga06} is not taken into account, however no
systematic error is ascribed due to that. An additional systematic
uncertainty reflects the precision of beam line and apparatus
description in the MC simulation.

{\bf Trigger efficiency} correction
$\epsilon(K_{e2})/\epsilon(K_{\mu2})=99.9\%$ accounts for the fact
that $K_{e2}$ and $K_{\mu2}$ decay modes are collected with
different trigger conditions: the $E>10$~GeV LKr energy deposition
signal enters the $K_{e2}$ trigger only. A conservative systematic
uncertainty of 0.3\% is ascribed due to effects of trigger dead time
which affect the two modes differently. {\bf LKr global readout
efficiency} $f_{\rm LKr}$ is measured directly to be
$(99.80\pm0.01)\%$ and stable in time using an independent LKr
readout system.

\section{Result and conclusions}

\begin{figure}[tb]
\begin{center}
{\resizebox*{0.5\textwidth}{!}{\includegraphics{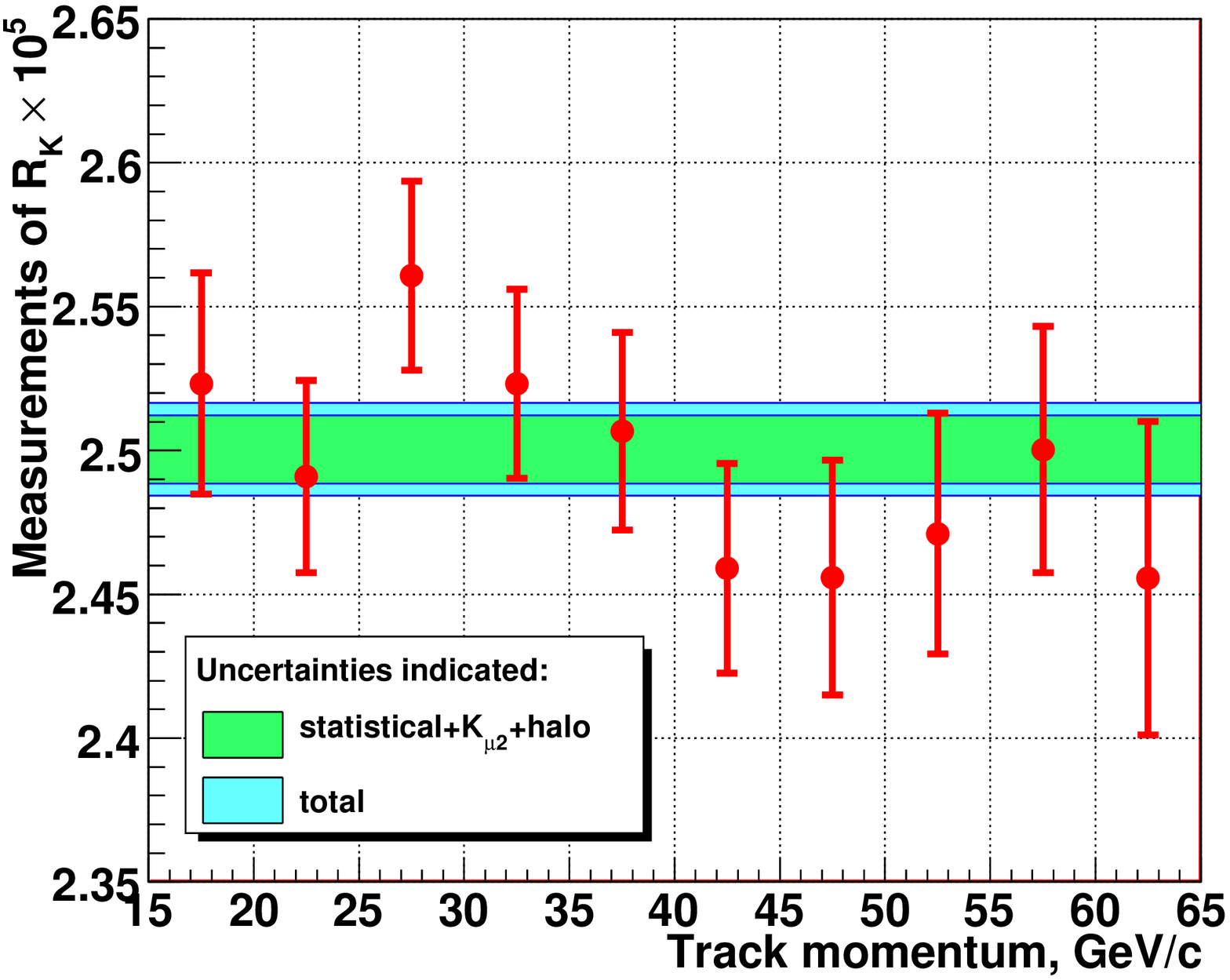}}}%
{\resizebox*{0.5\textwidth}{!}{\includegraphics{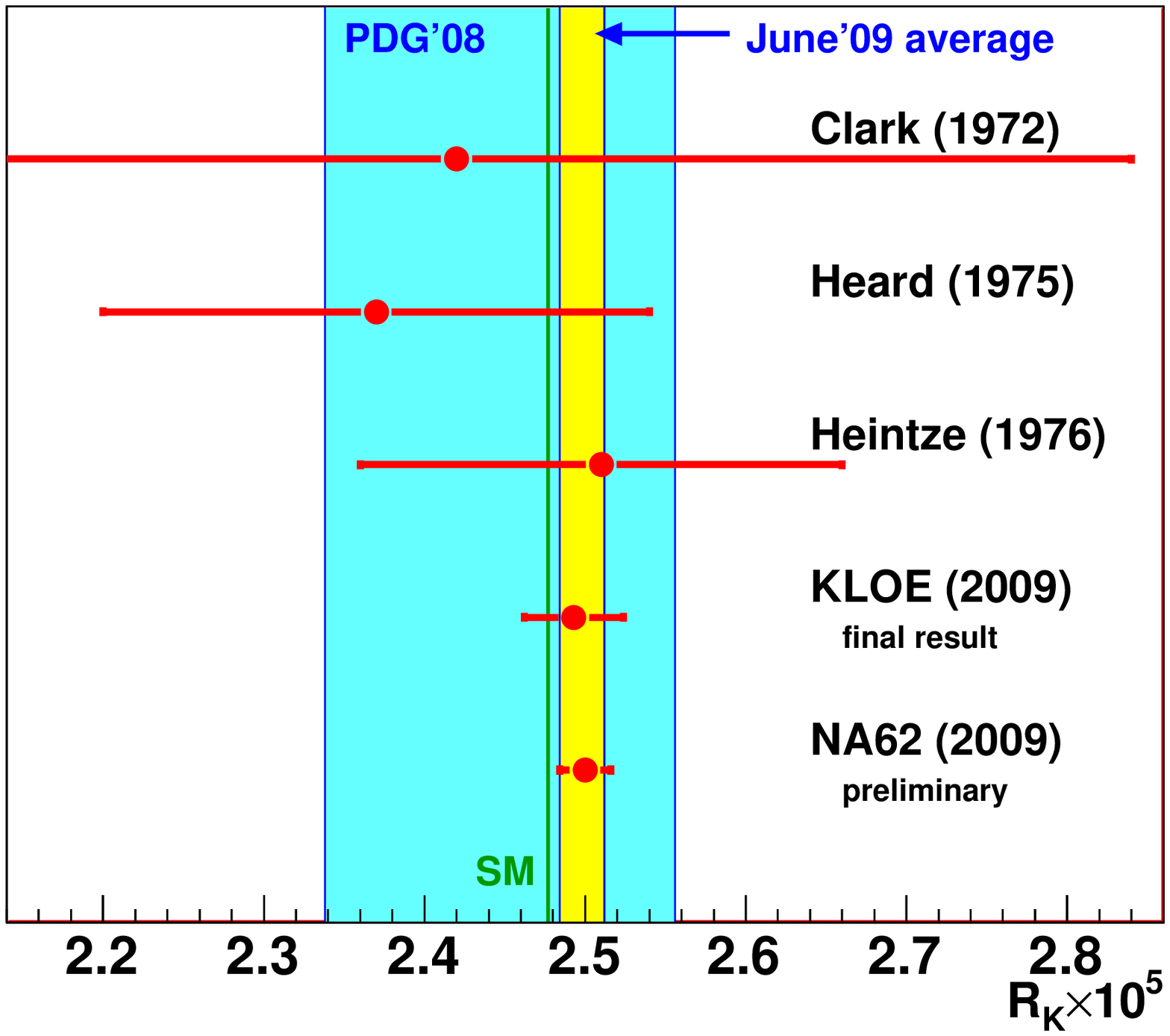}}}
\end{center}
\vspace{-3mm} \caption{(a) Measurements of $R_K$ in lepton momentum
bins. (b) The world average of $R_K$.} \label{fig:rk}
\end{figure}

The independent measurements of $R_K$ in lepton momentum bins, and
the result combined over the momentum bins are presented in
Fig.~\ref{fig:rk}a. The uncertainties of the combined $R_K$ are
summarised in Table~\ref{tab:err}. The preliminary result is
$R_K=(2.500\pm0.012_{\rm stat.}\pm0.011_{\rm syst.})\times 10^{-5} =
(2.500\pm0.016)\times 10^{-5}$, which is consistent with the SM
expectation. Analysis of the whole 2007--08 data sample is expected
to decrease the uncertainty of $R_K$ down to 0.4\%. A summary of
$R_K$ measurements is presented in Fig.~\ref{fig:rk}b: the current
world average is $(2.498\pm0.014)\times 10^{-5}$.

\begin{table}
\begin{center}
\begin{tabular}{lc|lc|lc}
\hline Source & $\delta R_K\times 10^5$ & Source & $\delta R_K\times 10^5$ & Source & $\delta R_K\times 10^5$\\
\hline
Statistical         & 0.012 & Beam halo      & 0.001 & Geom. acceptance  & 0.002\\
$K_{\mu2}$          & 0.004 & Positron ID    & 0.001 & Trigger dead time & 0.007\\
$K_{e2\gamma}$ (SD) & 0.004 & IB simulation  & 0.007\\
\hline
\end{tabular}
\end{center}
\caption{Summary of uncertainties of $R_K$: statistical and
systematic contributions.} \label{tab:err}
\end{table}



\section*{References}
\begin{thebibliography}{99}
%
\bibitem{ho93}
W.S. Hou, Phys. Rev. {\bf D48} (1993) 2342.
%
\bibitem{is06}
G. Isidori and P. Paradisi, Phys. Lett. {\bf B639} (2006) 499.
%
\bibitem{ci07}
V. Cirigliano and I. Rosell, Phys. Rev. Lett. {\bf 99} (2007)
231801.
%
\bibitem{ma06}
A. Masiero, P. Paradisi and R. Petronzio, Phys. Rev. {\bf D74}
(2006) 011701.
%
\bibitem{ma08}
A. Masiero, P. Paradisi and R. Petronzio, JHEP {\bf 0811} (2008) 42.
%
\bibitem{el09}
J. Ellis, S. Lola and M. Raidal, Nucl. Phys. {\bf B812} (2009) 128.
%
\bibitem{pdg}
C. Amsler {\it et al.} (PDG), Phys. Lett. {\bf B667} (2008) 1.
%
\bibitem{am09}
F. Ambrosino {\it et al.}, Eur. Phys. J. {\bf C64} (2009) 627.
Erratum-ibid. {\bf C65} (2010) 703.
%
\bibitem{fa07}
V. Fanti {\it et al.}, Nucl. Instrum. Methods {\bf A574} (2007) 433.
%
\bibitem{ga06}
C. Gatti, Eur. Phys. J. {\bf C45} (2006) 417.
%
\bibitem{ke97}
S.R. Kelner, R.P. Kokoulin and A.A. Petrukhin, Phys. Atom. Nucl.
{\bf 60} (1997) 576.
%
\bibitem{mi50}
L. Michel, Proc. Phys. Soc. {\bf A63} (1950) 514.
%
\end{thebibliography}
\end{document}